# Nonlinear optical response from excitons in soliton lattice systems

Kikuo Harigaya

*Physical Science Division, Electrotechnical Laboratory, Umezono, Tsukuba, Ibaraki 305, Japan*

**Abstract**

Exciton effects on conjugated polymers and the resulting optical nonlinearities are investigated in soliton lattice systems. The magnitude of the THG at the 10 percent doping increases by the factor about $10^2$ from that of the neutral system. The huge increase by the order two is common for several choices of Coulomb interaction strengths, and is seen in open systems as well as in periodic systems. We also find that the THG enhances much larger when the number of solitons is two in the open boundaries. This new fact should be attractive to the experimentalists.

*Keywords:* Semi-empirical models and model calculations, Electron density and excitation spectra calculations, Many-body and quasiparticle theories, Low-bandgap conjugated polymers

## 1. Introduction

The doping effects in conjugated polymers and their linear and nonlinear optical responses are fascinating research topics because of their importance in the scientific interests as well as technological developments. In the previous papers [1,2], we have theoretically considered exciton effects in the soliton lattice states of doped systems, and have investigated the nonlinear optical response properties of the periodic systems. However, all of the polymers are not regarded as long enough, and chain ends might effect on their excitation properties and nonlinear optical responses. In this paper, we show that the huge increase of the THG upon doping is seen in open systems, too. The effects of the chain ends are stronger when the number of the solitons is two. This is due to the trapping of the solitons near to the chain ends.

We use the SSH Hamiltonian [3] with the Ohno-type Coulomb interactions [4] among electons, $W(r) = 1/\sqrt{(1/U)^2 + (r/aV)^2}$, where the quantity $W(0) = U$ is the strength of the onsite interaction, $V$ means the strength of the long range part, $r$ is the distance between lattice sites, and $a$ is the mean bond length. The model is treated by the method used in [1,2]. In order to demonstrate the magnitude of the third harmonic generation (THG), we use the number density of the CH unit, which is taken from *trans*-polyacetylene: $5.24 \times 10^{22} \mathrm{cm}^{-3}$ [5]. We also use the intersite transfer integral $t = 1.8 \mathrm{eV}$ in order to look at numerical data in the esu unit.

## 2. THG of soliton lattice systems

First, we calculate the optical spectra and consider exciton effects. Figure 1(a) shows the typical optical absorption spectrum at the 2% soliton concentration for $(U,V) = (4t, 2t)$. There are two main features around the energies $0.7t$ and $1.4t$. The former originates from the excitons where an electron is excited from the soliton band to the conduc-tion band, and the latter comes from the excitons where an electron is excited between the continuum states.

Figure 1(b) displays the absolute value of the THG against the excitation energy $\omega$. The abscissa is scaled by the factor 3 so that the features in the THG locate at the similar points in the abscissa of Fig. 1(a). The large feature at about $\omega = 0.22t$ comes from the lowest excitation from the soliton band to the conduction band and the other feature at about $\omega = 0.26t$ comes from the higher excitations. The features due to excitons between continuum states extend from $\omega = 0.48t$ to the higher energies. The exciton between the soliton band and the conduction band gives rise to the large optical nonlinearity.

Next, Fig. 2 displays the variations of the absolute value of $\chi^{(3)}_{\mathrm{THG}}(0)$ for $(U,V) = (4t, 2t)$. The plots are the numerical data: the open symbols are for the periodic systems (already reported in [2]), and the closed symbols are for the present calculations of the systems with chain ends. The dashed lines are the guide for eyes for the plots of the periodic systems, showing the overall behavior for each system size $N$. The off-resonant THG near zero concentration increases very rapidly, but the THG is still increasing for a few percent to 10% soliton concentration. The increase between the zero concentration and the 10% concentration is by the factor about 100. The behavior is common for the two boundary conditions.

When we look at the enhancement factor in detail, it is found that the enhancement is particularly large for the systems with two solitons. This is clearly seen when we take the ratio of the THG value of the open system with respect to that of the periodic system. The calculations have been done for $(U,V) = (4t, 2t)$, and the ratio is shown in Fig. 3. Except for the three plots near the concentration 2%, the ratio is almost constant and about 3. It becomes more than 10 for the two soliton systems. Thus, we have found that the case with two solitons is special mainly owing to the chain end effects. It should be stressed that the large THG due to the presence of chain ends could be used as a tool for increasing nonlinear optical responses experimentally. In two soliton solutions with the open boundary condition, the inter-soliton distance becomes longer than in the system with periodic boundaries [6]. This gives rise to the larger expectation values of the dipole moment operators, and thus we obtain the huge enhanced THG in the system with two solitons.

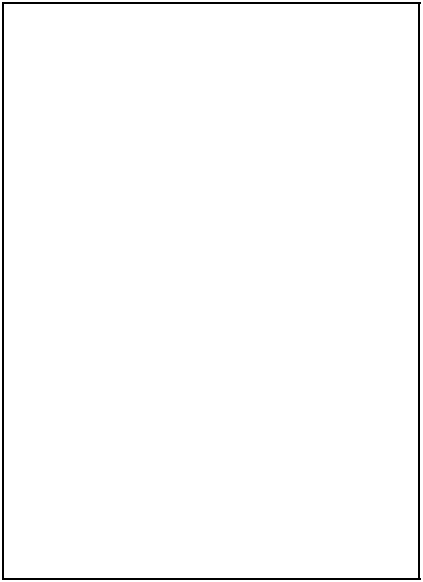

Fig. 1. (a) The optical absorption spectrum and (b) the absolute value of the THG. The broadening $\gamma = 0.05t$ is used in (a), and $\eta = 0.02t$ is used in (b). The absorption is shown in the arbitrary units, and the nonlinear optical response is in the esu unit.

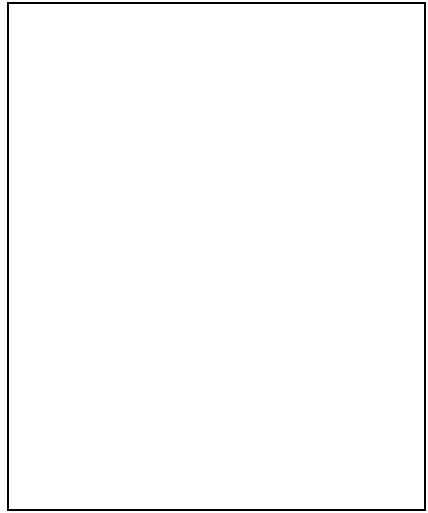

Fig. 2. The absolute value of the THG at $\omega = 0$ v.s. the soliton concentration for $(U, V) = (4t, 2t)$. The numerical data are shown by the triangles ($N = 80$), circles ($N = 100$), and squares ($N = 120$), respectively. The data of the system with the periodic boundary condition are shown by the open symbols. The data with the open boundaries are shown by the closed symbols.

## 3. Summary

We have considered the off-resonant nonlinear susceptibility as a guideline of the strength of the nonlinearity in the doped conjugated polymers. The off-resonant THG has been calculated with changing the system size and the soliton concentration for the chains with open boundaries. It has been shown that the magnitude of the THG at the 10 percent doping increases by the factor about $10^2$ from that of the neutral system. The huge increase by the order two is common for the several choices of Coulomb interaction strengths, and is seen in the open systems as well as in the periodic systems.

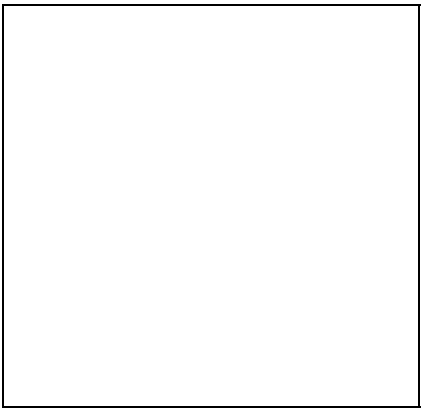

Fig. 3. The ratio of the absolute values of the THG between the open and periodic boundary conditions, shown against the soliton concentration. The numerical data are shown by the triangles ($N = 80$), circles ($N = 100$), and squares ($N = 120$), respectively.